\documentclass[aps,showpacs,twocolumn,floatfix,pre,superscriptaddress]{revtex4-1}
\usepackage{graphicx}
\usepackage{amsfonts}
\usepackage{amsmath}
\usepackage{amssymb}
\usepackage{multirow}
\usepackage[usenames,dvipsnames]{color}

\usepackage{bm}

\usepackage{upgreek}

\begin{document}

\title{L\'{e}vy walks with velocity fluctuations}
\date{\today}
\author{S. Denisov}
\affiliation{Institute of Physics, University of
Augsburg, Universit\"{a}tstrasse~1, D-86159 Augsburg, Germany}
\author{V. Zaburdaev}
\affiliation{School of Engineering and Applied Sciences, Harvard University, 29 Oxford street, Cambridge, MA 02138, USA}
\author{P. H\"{a}nggi}
\affiliation{Institute of Physics, University of
Augsburg, Universit\"{a}tstrasse~1, D-86159 Augsburg, Germany}

%\author{S. Denisov$^1$, V. Zaburdaev$^2$,  and P. H\"{a}nggi$^1$}
%\address{$^1$ Institute of Physics, University of
%Augsburg, Universit\"{a}tstrasse~1, D-86159 Augsburg, Germany}
%\address{$^2$ School of Engineering and Applied Science, Harvard University, 29 Oxford street,Cambridge, MA 02138, USA}

\begin{abstract}
The standard L\'{e}vy walk is performed by a particle that moves ballistically between randomly occurring  collisions, when the intercollision time is a random variable governed by a power-law distribution. During instantaneous collision events the particle  randomly changes the direction of motion but maintains the same constant speed. We generalize the standard model to incorporate velocity fluctuations into the process. Two types of models are considered, namely,  (i) with a  walker changing the direction and absolute value of its velocity during collisions only, and (ii) with a walker whose velocity continuously fluctuates. We present full analytic evaluation of both models and
emphasize the importance of initial conditions. We show that the type of the underlying  L\'{e}vy walk process can be 
identified by looking at the ballistic regions of the diffusion profiles. Our analytical results are corroborated 
by numerical simulations.
\end{abstract}

%\submitto{Journal of Physics A: Mathematical and Theoretical}
\maketitle

\section{Introduction}   The concept of random walk is one of the cornerstones of statistical physics \cite{book1, book2}. It is a universal  toolbox that can be used to study the dynamics of almost any physical system. Namely, the time evolution of the system  can be represented by a trajectory, $\mathbf{r}(t)$, in the corresponding coordinate or state space. Then, the system dynamics can be quantified by a mean square displacement (MSD), $\sigma(t) = \langle \mathbf{r}^2(t) \rangle$. Whereas in the  case of  the standard Brownian dynamics  the corresponding MSD scales linearly in time, $\sigma(t)  \sim t$, the hallmark of many complex systems,  \textit{anomalous diffusion}, is characterized by a nonlinear time dependence of the MSD,  $\sigma(t) \sim t^\alpha$, with $\alpha \neq 1$ \cite{klafter}. 

The phenomenon of anomalous diffusion is pertinent to the processes whose dynamics is dominated by long time or/and space correlations \cite{hanggi1}. The case of superdiffusion,  $\alpha > 1$, constitutes an intriguing limit. Real-life superdiffusive dynamics implies that  a walker -- an atom in an optical lattice \cite{atom}, a tracer in a turbulent flow \cite{tracer}, a predator hunting for food \cite{predator}, or a mussel in a bunch of peers \cite{mussel} -- explores its environment much faster then its Brownian 'colleagues' while still moving with a  bounded velocity, $|\mathbf{v}| < v_{\text{max}}$. The corresponding space-time dynamics is constrained to a cone, so that at a given time $t$ the walker is always located within the space region $\left|\mathbf{r}(t)-\mathbf{r}(0)\right|\leq v _{\text{max}}  t$.

The L\'{e}vy-walk (LW) process \cite{LW} represents a simplest stochastic model which combines both key ingredients, that are the superdiffuse evolution and the finiteness of the velocity of motion. In one-dimensional case the standard LW process can be sketched as follows: a walker performs a sequence of mutually uncorrelated ballistic flights of random duration but of fixed velocity, $v_0$. The flights are separated by the instantaneous collisions at which the walker changes the direction of its motion, taking randomly  either negative or positive direction. The time between consecutive collisions, and, therefore, the duration of  a single-flight event, is distributed according to a probability density function (PDF), $\psi(\tau)$, with a power-law asymptotic
\begin{equation}
\psi(\tau) \propto (\tau/\tau_{0})^{-\gamma-1},
\label{eq:psi}
\end{equation}
where a constant $\tau_{0}$ sets the characteristic timescale.  The exponent $\gamma$ in Eq.(\ref{eq:psi}) determines explicitly the scaling of the corresponding  MSD. Namely, $\alpha = 1$ when $\gamma>2$ (normal diffusion), $\alpha = 3-\gamma$ when $1<\gamma<2$ (superdiffusion),  and the choice of the exponent from the interval $0<\gamma<1$ leads to the ballistic diffusion, $\alpha = 2$ \cite{LW}. If an ensemble of particles, initially localized at $x = 0$, starts to spread at time $t = 0$, the corresponding propagators, i. e.  the PDFs to find a particle at a point $x$ at a time $t$, are all restricted  to the  cone $[-v _{0} t, v_{0} t]$, but have  different  shapes -- Gaussian profiles in the case of normal diffusion,  profiles in a form of  L\'{e}vy distributions in the case of superdiffusion, and  U-shaped profiles in the  ballistic limit \cite{scalingpaper}. Note, however, that all three propagators  exhibit a sharp cutoff at the points $\arrowvert x \arrowvert = v_0t$.

In addition to the  examples already listed, the LW formalism has found other successful applications, such as the description of DNA nucleotide patterns \cite{dna}, modeling the dynamics of an ion placed into an optical lattice \cite{ion}, analysis of the evolution of magnetic holes in ferrofluids \cite{ferro},  and engineering of L\'{e}vy glasses \cite{glass}. The LW ideology is flexible and leaves room for potential generalizations and modifications thus allowing to construct and tailor models which are able to mimic real processes in greater detail. For example, the condition of the constant velocity of ballistic motion is barely the case in real systems -- neither a foraging deer \cite{deer} nor an ion  moving through the optical lattice \cite{ion} are subjected to this strict condition. It is therefore worthwhile to construct extensions of the standard LW model that are able to take into account the effects of velocity fluctuations. 

Although it is intuitive that the light-cone cutoff  will be smeared by velocity variations, no further insight can be achieved without specification of the statistical properties of velocity fluctuations and their generating mechanisms. In this paper, we develop two generalizations of the standard one-dimensional LW scheme  which include velocity alternations. For the same flight-time PDF, Eq.(\ref{eq:psi}), both models yield the same MSD exponent and produce identical profiles of propagators in the \textit{innermost region} of the propagation cone. Of special importance are the \textit{ballistic regions}, where the shape of propagators exhibits a model-specific behavior. 

The paper is structured as follows. In Sec. II we define the models. In Sec. III we construct the corresponding transport equation for propagators and discuss the role of initial conditions.  Section IV is devoted to the asymptotic analysis of propagator profiles, especially at the regions of ballistic fronts. Section VI summarizes the results of the paper and discusses  possible applications of the proposed formalism.

\section{Models}

\subsection{Model A: L\'{e}vy walk with alternating velocities}  Consider a LW process where a particle performs every ballistic flight with constant speed  but randomly changes it during  scattering events between the flights. Velocity $v$ is now a random variable governed by a certain PDF, $h(v)$. A trivial bimodal PDF of a form 
\begin{equation}
h(v) = [\delta(v-v_0) + \delta(v+v_0)]/2 
\label{eq:pdf_v}
\end{equation}
corresponds to the standard LW scheme \cite{LW}. Velocity dynamics introduces an additional degree of freedom to the process, so that the resulting type of diffusion is determined now both by the flight-time PDF, $\psi(\tau)$, see Eq.(\ref{eq:psi}), and the velocity distribution, $h(v)$.   Here we consider a particular case when the delta-function in Eq.(\ref{eq:pdf_v}) is replaced by a hump-like distribution of finite variance, $\delta(v) \rightarrow \varDelta(v)$, %$\int_{-\infty}^{\infty} x \varDelta(x) dx= 0$, 
$\Sigma^2_{A}=\int_{-\infty}^{\infty} v^2 \varDelta(v) dv< \infty$. Since this work deals with \textit{velocity fluctuations}, below we will assume that the variance of humps is small compared to the average speed, $\Sigma_{A}\ll v_0$.
%In the context of the current work it is important to remark that the such a distribution of velocities either completely eliminates %the existence of the light front, or makes the boundary of the front not sharp. 

\subsection{Model B: L\'{e}vy walk with fluctuating velocity}
Consider a random walk process, where the velocity of a walker is not constant during the single flight event but fluctuates around an average value, $v_0$.  Phenomenologically, these fluctuations can be attributed either to some internal mechanisms -- chaotic precession of the magnetic moment of a ferrofluid particle that modifies the interaction of the particle with an external magnetic field \cite{ferro2}, or complex neural processes in the brain of foraging bumblebee \cite{deer}, or to some external mechanisms, like interaction of a moving nano-colloidal particle with an active medium \cite{kapral}. A particular variant of this  model has been introduced in Ref. \cite{prl}, where it has been used to describe the perturbation spreading in one-dimensional many-particle systems. 

The dynamics of the particle during a single flight event can be described by a Langevin equation,
\begin{equation}
\dot{x}=v_0+\xi(t),
\label{LE}
\end{equation}
where $\xi(t)$ is a Gaussian delta-correlated  noise with zero-mean and finite variance, $\left<\xi(t),\xi(t')\right>=D_v\delta(t-t')$, $D_{v}>0$. By integrating the above equation over some interval of time $\tau$, we obtain:
\begin{equation}
x(t+\tau)=x(t)+v_0\tau+w(\tau)
\end{equation}
The new stochastic variable, $w(\tau)=\int_{t}^{t+\tau}\xi(t')dt'$, can be characterized by a PDF $p(w,\tau)$, which is a Gaussian distribution with a dispersion $\Sigma_{B}^2=\left<(x-v_0t)^2\right>=D_v\tau$. Therefore, if a particle starts its flight of duration $\tau$ at a point $x$ with a velocity $v_0$, it will arrive at the point $x+v_0\tau+w(\tau)$, where $w$ is a random variable with the PDF $p(w,\tau)$.

Having  the microscopic descriptions of both models, we can now derive the evolution equations for the corresponding propagators, $P(x,t)$.

\section{Evolution equations for propagators}

We start with derivation of  an evolution equation for the propagator of a combined model, $A \otimes B$. The corresponding process is generated by a walker  that chooses its velocity from a distribution $h(v)$ at the beginning of the $i$-th flight, $v_i$, and then moves unidirectionally, with the instantaneous velocity, $\tilde{v}_i(t)$,  fluctuating around $v_i$. The velocity fluctuations are characterized by an universal PDF, $p(w,t)$.  Transport equations for a model, $A$ or $B$, can be obtained as particular cases either by setting  $D_v = 0$ (model $A$) or  assuming $h(v)=[\delta(v-v_0)+\delta(v+v_0)]/2$ (model $B$).

We follow the standard procedure \cite{derivation, zab2}, and introduce a space-time PDF for the collision events, $\nu(x,t)$, which gives the probability to observe collision in a point $x$ at a time $t$. It satisfies the following balance equation:
\begin{widetext}
\begin{equation}
\nu(x,t)=\int\limits_{-\infty}^{\infty}\int\limits_{-\infty}^{\infty}dvdw\int\limits_{0}^{t}\nu(x-v\tau-w,t-\tau)\psi(\tau)h(v)p(w,\tau)d\tau
+\varphi(t)\!\!\!\!\int\limits_{-\infty}^{\infty}h(v)p(x-vt,t)dv\label{eq:nu}.
\end{equation}
\end{widetext}
Here we assumed that all particles were launched from the point $x=0$, i.e. $P(x, t = 0) = \delta(x)$. 
The first summand on the right hand side of Eq.(\ref{eq:nu}) accounts for the particles that had changed the direction of their flights before the observation time $t$, while the second term  accounts for the particles that were flying during the whole observation time. If a particle starts at $x=0$ with a certain velocity $v$, the position of the first scattering event is influenced by the velocity fluctuations and is given by the PDF  $\int_{\infty}^{\infty}\delta(x-vt-w)p(w,t)dw = p(x-vt,t)$. %, while the function  $\psi(t)$ gives the probability to finish the first flight at time $t$. 
The prefactor of the second integral in Eq.(\ref{eq:nu}), $\varphi(t)$, defines the PDF of having the first scattering event at time $t$. We shall specify the exact functional form of $\varphi(t)$ in the next subsection, when addressing two different types of initial conditions. 

The PDF $\nu(x,t)$ allows us to define the corresponding propagator,   
\begin{widetext}
\begin{equation}
 P(x,t)=\!\!\!\!\int\limits_{-\infty}^{\infty}\int\limits_{-\infty}^{\infty}dvdw\int\limits_{0}^{t}\nu(x-v\tau-w,t-\tau)\Psi(\tau)h(v)p(w,\tau)d\tau
+\Phi(t)\!\!\!\!\int\limits_{-\infty}^{\infty}h(v)p(x-vt,t)dv.
\label{eq:n}
\end{equation}
\end{widetext} 
Here $\Psi(t)$ is the probability to continue the flight  that started at $\tau=0$ up to time $t$, % provided that at the beggining of the observation the particle was in the state of fligth,
\begin{equation}
\Psi(t)=1-\int_{0}^{t}\psi(\tau)d\tau. 
\end{equation}
The second summand on the right hand side of Eq.(\ref{eq:n}) is weighted with the function $\Phi(t)$, which is the probability to continue the very first flight during the whole observation time $t$.  Both functions, $\varphi(t)$ and $\Phi(t)$,  explicitly depend on the type of initial conditions that we are going to discuss in the following subsection.

%In both equations (\ref{eq:nu}, \ref{eq:n}) convolution type integrals are averaged over all possible combinations of flight times and velocities, as well as realizations of fluctuations.

\subsection{Equilibrated vs. non-equilibrated initial conditions}
There are two types of initial conditions which are  relevant in the physical context  \cite{tunaley}.

The first type, the so-called \textit{non-equilibrated} initial condition, assumes that all particles started their first flights at $t = 0$. In this case we have
\begin{eqnarray}
\varphi(t) \equiv \psi(t), \label{eq:noneq1}\\ 
\Phi(t)\equiv \Psi(t).
\label{eq:noneq2}
\end{eqnarray}

The second type, the \textit{equilibrated} initial conditions, corresponds to the situation when  all particles are already  in the stationary regime at time $t=0$. That means that every particle was initially in the state of flight with probability one, and all of them have long (possibly infinite) 'walking' histories. The following thought experiment might help to understand the equilibrated setup better. Assume that an ensemble of  particles was created at  $t=-t_1$. At the time $t=0$ we take the actual position of every particle as  a reference point, from which we count the displacement of the particle for time $t > 0$. It may be considered as though at initial time $t=0$ we instantaneously tagged all particles to the point $x=0$, without affecting their performance. The limit $t_1 \rightarrow \infty$ corresponds to the equilibrated setup. 

Consider now Eq.(\ref{eq:nu}) for $P(x, t=0) = \delta(x)$ without velocity fluctuations and in the non-equilibrated setup, meaning that $\varphi(t)=\psi(t)$. %Consider next Eq.(\ref{eq:nu}) without velocity fluctuations. 
This so simplified integral equation can  be solved by applying the Laplace transform with respect to time $t$. Convolution integrals are rendered as algebraic products in the Laplace space and, therefore, the frequency of velocity changes is given by the following expression,
\begin{equation}
\overline{\nu}(s)=\frac{P_0\overline{\psi}(s)}{1-\overline{\psi}(s)}.
\label{eq:nu_laplace}
\end{equation}
Here,  an overline denotes the Laplace transform and $s$ is the coordinate in the Laplace space. In general, the PDF for a particle to experience a collision after the start of observation depends on how long a timespan this particle was flying before. Therefore, for any particle at  a time $t$, we like to know its flight time. The corresponding PDF, $N(t,\tau)$, shows how many particles at the given point in time have a flight time $\tau$:
\begin{equation}
N(t,\tau)=\Psi(\tau)\nu(t-\tau)+\Psi(\tau)\delta(t-\tau).
\label{eq:Ntau}
\end{equation}
Again, with the help of the Laplace transform and using its shift property we find:
\begin{equation}
\overline{N}(\tau, s)=\frac{\Psi(\tau)\text{e}^{-s\tau}}{1-\overline{\psi}(s)}.
\label{eq:Ntau_laplace}
\end{equation}
Consider now the limit of large times, that in the Laplace space corresponds to small values of $s$. If the mean flight time exists, i.e., $\left<\tau\right>=\int_{0}^{\infty}\tau\psi(\tau)d\tau<\infty$, which is always the case for $\gamma > 1$, the leading terms in the expansion of the Laplace transform $\overline{\psi}(s)$ with respect to small $s$ can be written as $\overline{\psi}(s)\simeq 1-s\left<\tau\right>$.  Substituting it into Eq. (\ref{eq:Ntau_laplace}) and inverting the Laplace transform we arrive at
\begin{equation}
N(t,\tau)=\frac{\Psi(\tau)}{\left<\tau\right>}.
\label{eq:Ntau_answer}
\end{equation}
This is yet another way to arrive at the central result of the renewal theorem \cite{Feller}, see also \cite{HausKerr, tunaley, sher}. Now, when we know the distribution of particles with respect to their flight times $\tau$, we can calculate the PDF of the first collision after the process has been initiated, $\varphi(t)$. By using the conditional probability formula we obtain:
\begin{equation}
\varphi(t)=\int\limits_{0}^{\infty}\frac{N(\tau)\psi(t+\tau)}{\Psi(\tau)}d\tau
\label{eq:phi}
\end{equation}
A similar expression can be written for the probability of having no collisions before time $t$, $\Phi(t)$. Substituting the distribution over the flight times, Eq. (\ref{eq:Ntau_answer}), in the above equation we arrive at:
\begin{eqnarray}
\varphi(t)&=&\left<\tau\right>^{-1}\int_{0}^{\infty}\psi(t+\tau)d\tau, \label{eq:equ1}\\
\Phi(t)&=&\left<\tau\right>^{-1}\int_{0}^{\infty}\Psi(t+\tau)d\tau.
\label{eq:equ2}
\end{eqnarray}

Different types of initial conditions naturally correspond to different experimental setups. If an ensemble of random walkers has been created and then immediately launched, then the non-equilibrated initial conditions is the proper setting. However, if the ensemble has already been evolved for a while, before one starts to measure ensemble characteristics, the equilibrated initial conditions would be the appropriate choice.

As it can been seen from  Eqs. (\ref{eq:noneq1}, \ref{eq:noneq2})  and Eqs.  (\ref{eq:equ1}, \ref{eq:equ2}), the case of equilibrated initial conditions is characterized by a more pronounced influence of the initial distribution of particles on the propagator evolution. We will unfold this important observation in the subsequent section.

\subsection{Solution of the evolution equations}

Equations (\ref{eq:nu}-\ref{eq:n}) can be studied further by using the combined Fourier and Laplace transforms in space and time domains correspondingly. Taking into account the shift property of the integral transform and convolution form of the integrals, the original equations (\ref{eq:nu}-\ref{eq:n}) can be reduced to the system of algebraic equations:
\begin{eqnarray}
\widetilde{\nu}(k,s)&=&\widetilde{\nu}(k,s)\overline{\left[\widehat{h}(k\tau)\widehat{p}(k,\tau)\psi(\tau)\right]}+\overline{\left[\widehat{h}(kt)\widehat{p}(k,t)\varphi(t)\right]}\nonumber\\
\widetilde{P}(k,s)&=&\widetilde{\nu}(k,s)\overline{\left[\widetilde{h}(k\tau)\widehat{p}(k,\tau)\Psi(\tau)\right]}+\overline{\left[\widehat{h}(kt)\widehat{p}(k,t)\Phi(t)\right]}\nonumber
\label{system}
\end{eqnarray}
We used overline and  hat notations to denote the Laplace and Fourier transforms, and the tilde notation for their combination, whereas $k$ ($s$) denote the coordinates in the corresponding Fourier (Laplace) space. The above system can be resolved straightforwardly, yielding,
\begin{eqnarray}
\widetilde{P}(k,s)&=&
\frac{\overline{\left[\Psi(\tau)\widehat{h}(k\tau)\widehat{p}(k,\tau)\right]}\overline{\left[\varphi(t)
\widehat{h}(kt)\widehat{p}(k,t)\right]}}{1-\overline{[\widehat{h}(k\tau)\widehat{p}(k,\tau)\psi(\tau)]}}\nonumber\\&+&\overline{\left[\widehat{h}(kt)\widehat{p}(k,t)\Phi(t)\right]} \;.\label{density_answ}
\end{eqnarray}

Note that no prior assumptions concerning the velocity, flight time or noise PDF's were made before to obtain the \textit{formal} solution in Eq. (\ref{density_answ}). 
An attribute 'formal', aside from denoting an exact character  of the final result, also carries a certain negative tone. Namely, it is practically impossible to handle such a complex expression as Eq.(\ref{density_answ}), especially when trying to map it backward onto the original space-time domain. There are only two possibilities left, namely,  either to resort to asymptotic analysis or to perform direct numerical calculations. We are going to pursue both options  in the section below.

\section{Asymptotic analysis}

The parameter space of the general model, given by  Eqs. (\ref{eq:nu}, \ref{eq:n}), is a highly-dimensional space and its detailed exploration is outside the scope of the present work. We remind that we are interested in the limit when velocity fluctuations are small compared to the characteristic velocity of walkers. For the model $A$ that assumes the limit $\Sigma_{A}\ll v_0$, while for the model $B$ it means that $\sqrt{D_\upsilon} \ll v_0$. It should be also noted that there are two types of contributions to the overall propagator. The first term on the right hand side of Eq.(\ref{eq:n})  describes the self-similar evolution of the particles that form the central part of the density profile, whereas the second term is a contribution which stems from the initial conditions and describes the behavior of the ballistic fronts. These contributions should be analyzed in more detail, and below we analyze them separately.

\subsection{Central part of the density distribution}
The essence of the asymptotic analysis routinely employed in random walks is to look at the large time and space limit, $x,t\rightarrow\infty$, which corresponds to the limit of small values of $k$ and $s$ in the Fourier/Laplace space, $k,s\rightarrow 0$ \cite{book2}. Therefore instead of using the full expressions for the corresponding transforms, only the leading terms in their expansions with respect to small $k$ and $s$ should be retained. It is possible to show that at the limit of small velocity fluctuations  the asymptotic behavior of the central part of the propagators for both models is identical to that of the standard  L\'{e}vy walk process \cite{LW}. In the case of L\'{e}vy walks, when $p(w,t)=\delta(w)$, the first term on the right hand side of the general expression given in Eq. (\ref{system}) can be  rewritten as: 
\begin{equation}
\widetilde{P}^{\text{LW}}\!=\!\frac{\left[\overline{\Psi}(s\!+\!ikv_0)\!\!+\!\!\overline{\Psi}(s\!-\!ikv_0)\right]\!\!\left[\overline{\varphi}(s\!+\!ikv_0)\!\!+\!\!\overline{\varphi}(s\!-\!ikv_0)\right]}{2-\left[\overline{\psi}(s+ikv_0)+\overline{\psi}(s-ikv_0)\right]}
\end{equation}
For the analysis of its asymptotic properties we make use of the expansion:
\begin{eqnarray}
\overline{\psi}(s &\pm& ikv_0)\simeq 1-\frac{\tau_0}{\gamma-1}(s\pm ikv_0)
-\Gamma[1-\gamma]t_0^{\gamma}(s\pm ikv_0)^{\gamma}\nonumber\\&+&\frac{\tau_0^2}{(\gamma-2)(\gamma-1)}(s\pm ikv_0)^{2}+O\left((s\pm ikv)^{1+\gamma}\right).\label{eq:expan}
\end{eqnarray}
Depending on $\gamma$ certain terms in the above expression take the leading role thus defining three major scaling regimes \cite{scalingpaper}. For $\gamma>2$ the mean square of the flight distance, $\left<(v_0\tau)^2\right>$, is finite, so that the corresponding transport process is the normal diffusion. In the intermediate regime, $1<\gamma<2$, the  mean squared flight distance diverges. In this regime the leading term of $k$ scales like $k^{\gamma}$ thus leading to anomalous diffusion  and L\'{e}vy-like profiles of the corresponding propagators. Finally,  long flights  dominate the process at the limit  $0<\gamma<1$, thus forming U-shaped propagators \cite{LW}. 

In the case of anomalous superdiffusion, $1<\gamma<2$, the Laplace/Fourier image of  the propagators  of both models, $A$ and $B$, in the limit of  small $k$ and $s$, is given by:
\begin{equation}
\widetilde{P}^{\text{LW}}\!\simeq\!\frac{1}{s+\tau_{0}^{\gamma-1}v_0^{\gamma}(\gamma-1)\Gamma[1-\gamma]k^{\gamma}\sin(\pi\gamma/2)}.
\end{equation}
In the space-time domain  it corresponds to the L\'{e}vy distribution with a characteristic power-law behavior of tails, exhibiting the following scaling properties \cite{LW, scalingpaper}:
\begin{equation}
P(x,t')\simeq \frac{1}{K u^{1/\gamma}} P \left(\frac{x}{K u^{1/\gamma}},t\right),~~~~ |x| \ll v_0 t \;,
\label{eq:scaling1}
\end{equation}
where $K \propto \tau_{0}^{1-1/\gamma}v_0$ and $u = t'/t$. 

Therefore, in the limit of small velocity fluctuations and for $1<\gamma<2$, the central part of the PDF is universal and is given by the  well-known L\'{e}wy walk propagator \cite{LW}.

\begin{figure}[t]
\center
\includegraphics[width=0.49\textwidth]{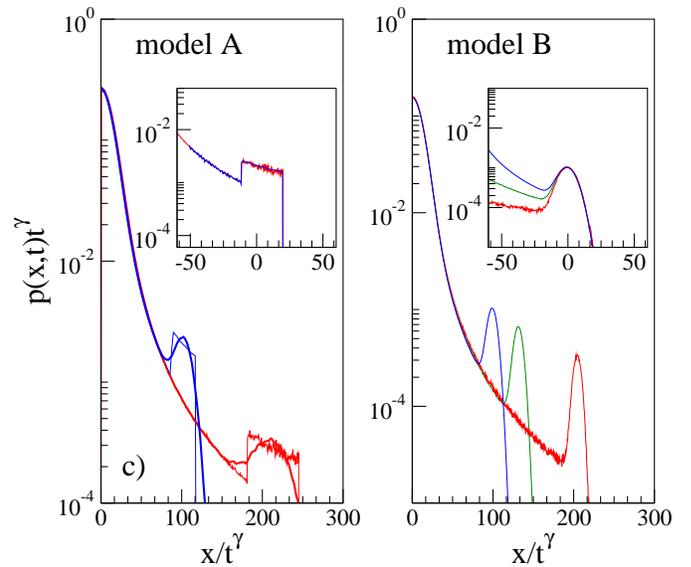}
\caption[]
{(color online) Rescaled propagators for generalized L\'{e}vy-walk processes with the exponent $\gamma = 5/3$;  at times $t=100$ and $600$ for model $A$ (left panel), and at $t = 100, 400,$ and $600$ for model $B$ (right panel) for equilibrated initial conditions. The profiles for the model $A$ are shown for two different velocity probability distribution functions, $\varDelta(v)$, rectangular and Gaussian, with the variance $\Sigma^2_{A} = 0.1$. Other parameters are $v_0 = 1$ and $D_\upsilon = 0.1$.
The insets show the ballistic front regions as a function of a shifted coordinate, $\bar{x}=x-v_0 t$, with a width and height rescaled by the corresponding power laws listed in Table 1. }
\label{fig3}
\end{figure}

\subsection{Ballistic humps}

The ballistic regions are model specific. Therefore, their analysis on the base of the combined model, introduced in Sec. III, is imposible and two models should be considered separately. 

Relations (\ref{eq:nu}-\ref{eq:n}) tell us that ballistic humps are formed by the particles which were flying from the very beginning, i.e.,  they either started their flight or they were already in the state of flying at the time $t=0$. Therefore the expression for the PDF in the ballistic hump regions reads
\begin{equation}
P_{\rm{hump}}(x,t)=\Phi(t)\int\limits_{-\infty}^{\infty}h(v)p(x-vt,t)dv.
\label{hump}
\end{equation}
Now the difference between two models, $A$ and $B$, is evident. In the case of random velocities, that is the case of  model $A$, the width of the ballistic humps is proportional to the observation time $t$ because the particles with slightly different initial velocities will separate ballistically, and the interparticle distances will grow linearly in time. The shape of a velocity PDF, $h(v)$, will be reproduced by the ballistic humps, see Fig. 1 (left panel).   In the case of delta-like velocity distribution $h(v)$, see Eq. (\ref{eq:pdf_v}), but with  velocity fluctuations, that is the case of model $B$, the particles flying from the very beginning will accumulate fluctuations during the spreading time, and, according to the central limit theorem, the width of the initially  $\delta$-like ballistic peaks will grow as $\sqrt{t}$. The shape of the peaks will be always Gaussian.  

The total number of particles in the hump, i.e., the area under the hump,  is governed by the survival probability $\Phi(t)$. This number is the same for both models and depends on the type of the initial conditions. Already now we can say that the height of the humps  decays faster for  model $A$, since its width increases faster. 
By substituting the power-law flight time PDF, Eq.(\ref{eq:psi}), into Eq. (\ref{hump}), we can explicitly calculate the scaling of the width, height, and total number of particles in the hump for two different models and two types of initial conditions. The result of the evaluation are presented with Table 1. %One can immediately  see that the humps in the case of  model A spread faster and also their height decays faster. 
For both models the equilibrated initial conditions lead to a slower decay of the hump height. On Fig. 1 we plot the propagators of both models for different times, where for the model $A$ we used two  velocity PDFs, $h(v)$,  with a Gaussian (thick solid lines) and a rectangular humps (thin lines) around characteristic velocity values $\pm v_0$, with $ v_0 = 1$. Remarkably, the rectangular shape of the velocity PDF is directly translated into the 
shape of ballistic fronts. In the case of Gaussian PDF, ballistic humps look similar for both models. However, the time evolution reveals the dramatic difference in the scalings of the humps' profiles, see Fig.1 and Table 1. 

%\begin{equation}
%P_{\text{hump}}(\bar{x},t') \simeq u^{-1/2}P_{\text{hump}}(\bar{x}/u^{\gamma-1/2},t),
%\label{eq:scaling2}
%\end{equation}
%where $u = t'/t$ and $\bar{x} = x - v_0 t$, see inset in Fig. 2. Note that this scaling  distinctly differs from the scaling  in  Eq. (\ref{eq:scaling1}).

%Therefore two models show different scalings of the ballistic humps, namely linear versus square root of the spreading time. This is a good  diagnostic tool which allows to determine the suitable model for the corresponding real process.

%\begin{table}%[htdp] 
%\caption{\label{table1}Scaling properties of ballistic humps for three different random walk models: Levy-walk, random walks with random velocities, fluctuating velocities, and equilibrated and non-equilibrated initial conditions in the superdiffusive ($1<\gamma<2$) regime.}
%%\caption{Scaling of humps}
%%\begin{center}
%
%\begin{tabular}{|c|c|c|c|c|}\hline
%\multicolumn{2}{|c|}{Model} & area & width & height \\ \hline
%\multirow{2}{*}{Levy walk} & non-equilibrated & $t^{-\gamma}$ & - & -\\
% & equilibrated & $t^{1-\gamma}$ & - & -\\ \hline
%\multirow{2}{*}{Random velocities} & non-equilibrated &  $ t^{-\gamma}$ & $ t$& $ t^{-1-\gamma}$\\
%& equilibrated &  $ t^{1-\gamma}$ & $ t$& $ t^{-\gamma}$\\ \hline
%\multirow{2}{*}{Fluctuating velocities}&non-equilibrated &  $ t^{-\gamma}$ & $ t^{1/2}$& $ t^{-1/2-\gamma}$\\
%&equilibrated &  $ t^{1-\gamma}$ & $ t^{1/2}$& $ t^{1/2-\gamma}$\\
%\hline
%\end{tabular}
%%\end{center}
%\end{table}

\begin{table}[htdp]
\caption{Scaling properties of ballistic humps of three different random walk models, standard L\'{e}vy-walk (LW), process with alternating ($A$) and  fluctuating  ($B$) velocities, for  equilibrated (eq.) and non-equilibrated (non-eq.) initial conditions. Here we address the regime of anomalous  superdiffusion, $1<\gamma<2$.}
\begin{center}
\begin{tabular}{|c|c|c|c|c|}\hline
\multirow{2}{*}{Model}& \multirow{2}{*}{Initial condition} & \multicolumn{3}{c|}{Scaling} \\ \cline{3-5}
& &width & height & area\\ \hline
\multirow{2}{*}{LW} & non-eq. & - & - & $t^{-\gamma}$ \\ %\cline{2-5}
& eq. & - & - &$t^{1-\gamma}$\\ \hline
\multirow{2}{*}{$A$} &non-eq.&  $ t$ & $ t^{-1-\gamma}$& $  t^{-\gamma}$ \\
&eq.&  $t$ & $ t^{-\gamma}$& $ t^{1-\gamma}$ \\ \hline
\multirow{2}{*}{$B$} &non-eq.&  $ t^{1/2}$ & $ t^{-1/2-\gamma}$& $ t^{-\gamma}$ \\
&eq.&  $ t^{1/2}$ & $ t^{1/2-\gamma}$& $ t^{1-\gamma}$ \\
\hline

\end{tabular}
\end{center}
\label{default}
\end{table}%

\section{Conclusions.}
In this paper we have presented two random walk models that describe stochastic transport phenomena with random velocities. Two different mechanisms of velocity fluctuations have been analyzed. These correspond to instantaneous alternations during collision events (model $A$) and continuous velocity fluctuations during the flights (model $B$). Both models can generate processes that exhibit anomalous superdiffusive behavior. However, corresponding diffusion profiles reveal essentially different behaviors in the ballistic regions, thus underlining the fact that ballistic fronts carry an important information about the origin of velocity fluctuations,  maintaining memory of the initial conditions. Therefore, ballistic humps may serve as a diagnostic tool which allows to calibrate velocity fluctuations and explore the internal dynamics of a random walker. Our analytical results  open the possibilities to study the evolution of complex systems, ranging from a bead moving in a colloidal medium to a motion of a bacterium, in which case velocity fluctuations are controlled by complex chemical circuits. Corresponding spatial-temporal patterns can be reproduced with relatively simple and transparent random walk models thus providing a new  tool for the analysis of these patterns.

%\acknoledgements
This work has been supported by the DFG Grant HA1517/31-2 (S.D. and P.H.) and DFG fellowship ZA593/2-1(V.Z.).

%\section{References}

\end{document}